\documentclass  {mn2e}

\newcommand\kms{$\rm{km\,s^{-1}}$}
\newcommand\HII{H\,{\sc ii}}

\input epsf
\epsfverbosetrue

\title[300.969+1.147 maser maps and magnetic field]
{Maser maps and magnetic field of OH 300.969+1.147}

\author[J. L. Caswell, B. Hutawarakorn Kramer, and J. E. Reynolds]
       {J. L. Caswell$^1$, B. Hutawarakorn Kramer$^{2,3}$, and J. E. 
Reynolds$^1$\\
$^1$ Australia Telescope National Facility, CSIRO, PO Box 76,
Epping, NSW 2121, Australia; james.caswell@csiro.au \\
$^2$ Max-Planck-Institut f\"ur Radioastronomie, Auf dem H\"ugel 69, 53121 
Bonn, Germany; bkramer@mpifr-bonn.mpg.de \\
$^3$ National Astronomical Research Institute of Thailand, Physics
Building, Chiang Mai University, Chiang Mai 50200, \\
Thailand; busaba@narit.or.th}

\date{Accepted .
      Received ;
      in original form 2009}

\pagerange{\pageref{firstpage}--\pageref{lastpage}}
\pubyear{2009}

\begin{document}

\maketitle

\label{firstpage}

\begin{abstract}

The southern maser site OH 300.969+1.147 has been studied using the Long 
Baseline Array of the Australia Telescope National Facility.  The 1665- 
and 1667-MHz hydroxyl ground-state transitions were observed 
simultaneously.  A series of maps with tenth-arcsec spatial resolution, at 
velocity spacing 0.09 \kms, and in both senses of circular polarization,
reveal 59 small diameter maser spots.  The spots are scattered over
2-arcsec, coincident with a strong ultracompact \HII\ region, at a 
distance of 4.3 kpc.
17 Zeeman pairs of oppositely polarized spots were found, all 
yielding magnetic field estimates towards us (negative), ranging from 
-1.1 to -4.7 mG, with a median value of -3.5 mG.  
Excited state masers of OH at 6035 MHz and 6030 MHz at this site also 
display Zeeman pairs revealing a magnetic field of -5.0 mG. 
Weak methanol maser emission is intermingled with the OH masers, but 
there is no detectable closely related water maser.  
The consistent magnetic field direction found within this site is a 
striking feature of several other maser sites associated with strong \HII\ 
regions studied in comparable detail.
We interpret the site as a mature region nearing the end of the brief 
evolutionary stage that can support maser emission.  

\end{abstract}

\begin{keywords}
masers - stars: formation - \HII\ regions - ISM: molecules - radio lines:
ISM.
\end{keywords}

\section{Introduction}

Molecular maser emission associated with newly formed massive stars 
commonly comprises many small spots spread over a site of typical extent 
30 milliparsec, enveloping the star.  
Hydroxyl masers are especially
useful probes of the immediate surroundings of the star since the Zeeman 
effect leads to a readily measurable separation of left and right 
circularly polarized emission in magnetic fields of a few mG.  
At most maser sites, it requires the high spatial resolution of Very Long 
Baseline Interferometry to unequivocally recognise individual Zeeman 
patterns, and to explore the homogeneity of the magnetic field within the 
masing region.  Pioneer observations began more than two decades ago, 
and there has now been increased recent activity from the VLBA, with 
new results for 18 targets (Fish, Reid, Argon \& Zheng 2005). The target 
properties are found to be diverse, but hint at clearly 
recognisable classes of object, demonstrating a need for similar studies 
of the numerous potential targets in the southern hemisphere.

	The southern hemisphere Long Baseline Array (LBA) of the Australia
Telescope National Facility (ATNF) allows high frequency resolution over a 
total bandwidth spanning both 
the 1665- and the 1667-MHz ground-state OH transitions.  Using three
stations, the spatial resolution of approximately 100 
milliarcsec (mas) is quite well suited to observations of masers at 
distances of a few kpc; it allows clear position discrimination between 
the numerous tight clusters of spots (of typical size $< 100$ mas) that 
are commonly spread over a total extent of an arcsec.  

Our program with the LBA to study the rich variety of 
masers in the southern sky began with OH 323.459-0.079 (Caswell \& 
Reynolds 2001);  the present study is similar, but improved by a lower 
noise level, and analysed at higher spectral resolution.

\section[]{Observations}

Observations were made over a 15h period 2000 August 17.  The array 
comprises the Parkes 64-m antenna, complemented by a 22-m paraboloid 
at Mopra, and the compact array of the Australia Telescope (ATCA) at 
Narrabri operating as a tied array (equivalent in sensitivity to a dish of 
46-m diameter).  
These yield baselines of approximately 119 km, 203 km and 321 
km, predominantly north-south.  The S2 tape system recorded both right- 
and left-hand senses of circular polarization (RHCP, LHCP) simultaneously,
with a bandpass centred at 1667 MHz, and limited to a 4-MHz bandwidth by a
digital filter with excellent flat response in amplitude and phase.  Our 
strategy was to observe a target for 25-min periods, bracketed by 5-min 
periods on 
a phase and secondary amplitude calibrator (1148-671, 1215-457 or 
1740-517, with total intensity flux densities of 1.78, 
3.37 and 7.47 Jy).  Flux densities are relative to the absolute 
calibration derived from 1934-638 (unpolarized, with total intensity of 
14.157 Jy).  
The target position of 300.969+1.147 (J2000 RA 12$^h$34$^m$53.20$^s$,
Dec. -61$^{\circ}$39${\arcmin}$40.0${\arcsec}$) 
was alternated with another target (337.505-0.053, to be discussed 
in a later paper) for the 
period when both were accessible to the Parkes telescope.  

\section[]{Data Reduction}

Each baseline required a unique correlator pass in order to achieve 
the high frequency resolution of 8192 channels across the 4-MHz band.  
The output from correlation of the signals from each polarization was 
binned into 5-sec integration periods.  
In the subsequent
processing using the AIPS reduction package, no Hanning smoothing was 
applied so that the final channel separation remains 0.48828125 kHz (= 
0.088 \kms), and the final velocity spectral resolution is 0.105 \kms\ 
(larger than the channel separation by a factor of 1.2 for uniform 
weighting).  

The calibrators were used to
derive first order phase and amplitude corrections to the target source
observations, and this calibration also established the correct phase
between RHCP and LHCP (and subsequently ensures precise registration of  
their relative positions).  The program `cvel' was used to align channels 
to the same velocity, and allow for the varying Doppler correction arising
from the Earth's motion during the observations, adopting rest frequency 
values of 1665.4018 and 1667.359 MHz.  After perusal of 
spectra from each polarization, strong
channels were selected, from which preliminary maps were made.

\begin{figure}
\centerline{\epsfxsize=8.4cm\epsfbox{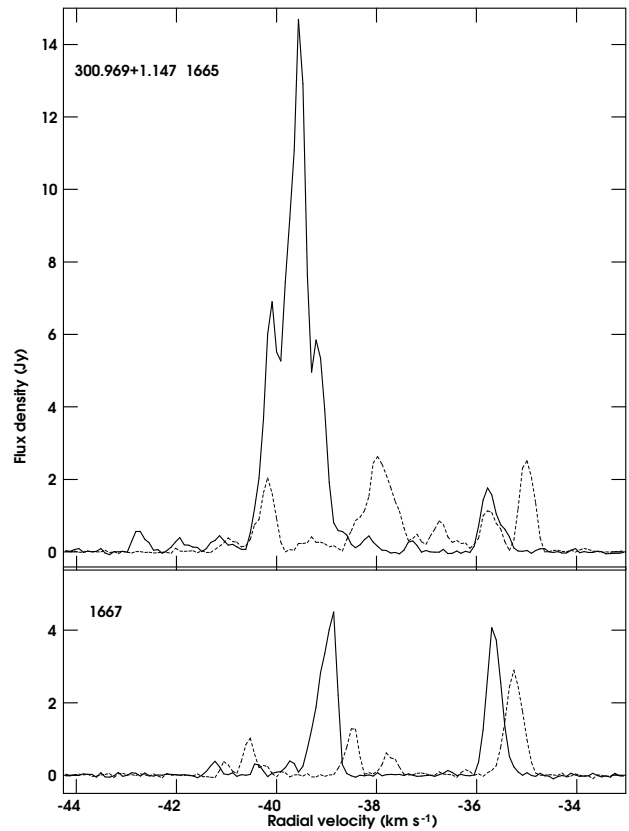}}
\caption{Spectra of the 1665- and 1667-MHz OH transitions towards
OH 300.969+1.147, taken 2000 Aug 17.  The two circular polarizations are
distinguished as RHCP (thick line) and LHCP (thin broken line).  The
intensities are scaled such that the total flux density is the sum (not
the average) of the two circular polarizations.  The spectra comprise
emission from many individual spots distributed over approximately 2
arcsec, as revealed by the LBA maps.}  
\end{figure}

Maps of the field using only the preliminary calibration did not allow the
measurement of an absolute position.  However the strongest 
(15-Jy) spectral feature, RHCP 1665 MHz at velocity -39.8 \kms, has a
known position derived from ATCA observations (Caswell 1998) of RA 
12$^h$34$^m$53.24$^s$, Dec. -61$^{\circ}$39${\arcmin}$40.3${\arcsec}$, 
with an uncertainty of 0.4 arcsec.  
It would be possible to use this feature as a phase reference to determine 
positions of all other features.  However, it turns out that a RHCP 
1667-MHz feature of somewhat lower amplitude, at velocity -35.8 \kms, is 
more suitable as a phase 
reference since it is more clearly distinct from other spectral 
features.  Relative to the ATCA measured feature, we show later from 
the current LBA observations that it is located at larger
RA by 0.03s (= 0.22 arcsec) and north by 0.67", ie at RA 
12$^h$34$^m$53.27$^s$, Dec. -61$^{\circ}$39${\arcmin}$39.63${\arcsec}$.  
This RHC 1667-MHz feature was then used as the reference feature for the 
maps.  The uncertainty in this absolute registration remains the same 
as the ATCA measurement, approximately 0.4 arcsec.

\begin{figure*}
\centerline{\epsfxsize=12cm\epsfbox{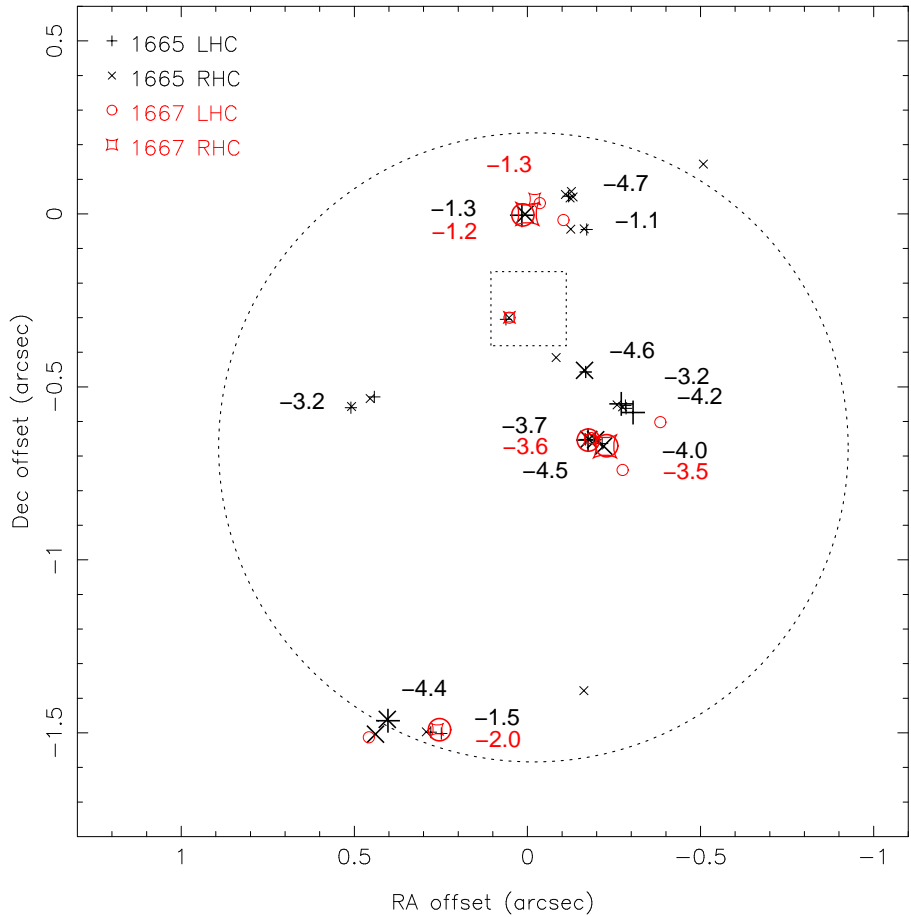}}
\caption{Spatial distribution of 59 maser spots in OH 300.969+1.147, with 
map reference position at RA 12$^h$34$^m$53.27$^s$, Dec. 
-61$^{\circ}$39${\arcmin}$39.63${\arcsec}$.   
The distributions of RHCP and LHCP for both the 1665 and 1667-MHz 
transitions are superposed and denoted by different-shaped symbols.  
Larger symbols denote 
features with peaks of 0.5 Jy or greater.  The Zeeman pairs are 
identified by the magnetic field value (mG) labelled near the 
corresponding pair.  The field derived at 1665 MHz is shown above the 
value derived at 1667 MHz when Zeeman pairs at the two transitions 
essentially coincide.   
See Table 1 for precise parameters of spectral features.  The 
approximate location of associated maser emission at 6035 MHz is denoted 
by a small dotted square.  The approximate boundary of an underlying \HII\ 
region is shown as a large dotted circle.} 
\end{figure*}

This final calibration of all data relative to a single RHCP 
1667-MHz feature reduces phase errors, establishes a correct 
absolute position, and maintains correct relative positions between 
polarizations and between transitions.  

In the velocity range -44 to -33 \kms, `cleaned' maps were made of each
spectral line channel (total 128 channels) for both transitions, and for 
both circular polarizations.  The cellsize used was 25 mas, and the 
restoring beamwidth was 72 x 130 mas (major axis at position angle 
94$^\circ$).  
The typical rms noise on a map was 8 mJy, allowing reliable detections of 
features down to 50 mJy, or weaker if present over several
channels.  At the few velocities where a strong feature exceeding 2 Jy is
present, the detection of weak features at some positions is limited by
dynamic range (owing to occasional 5 per cent sidelobe levels).

\section[]{Results}

\subsection{OH Spectra, maser spot positions and morphology}

OH spectra in Fig. 1 show the total maser emission from the site
at 1665 and 1667 MHz using the present observations.  
Earlier spectra, recorded 1982 February, can be seen in Caswell \& 
Haynes (1987a), and are similar to those measured when it was discovered 
(Robinson et al. 1974).  The 
present spectra from 2000 August (Fig. 1) resemble the earlier spectra, 
but have higher sensitivity.  Subsequent spectra from the Parkes telescope 
in 2004 November and 2005 October again resemble earlier spectra.  
Measurements of weak linear polarization in these new spectra will be 
discussed later insofar as they assist interpretation of the LBA maps 
which were not analysed for linear polarization.  

From the `cubes' of the images of RHCP and LHCP at 1665 and 1667 MHz, 
all emission features were measured using the `imfit' task in AIPS to 
derive 
positions and flux densities, and all appear to be of small angular size 
($< 25$ mas) relative to the beamwidth of 72 x 130 mas.

The positions measured for 59 individual features, corresponding to 
maser `spots', are listed in Table 1; they are spread over an elliptical 
region elongated 1.66 arcsec north-south.  Each row of Table 1 lists a 
narrow spectral feature (typically 0.4 \kms\ to halfpower), in a
single sense of circular polarization, at one of the two OH transitions of
1665 or 1667 MHz.  
The maser features are grouped by transition and sense of circular 
polarization, and then ordered by radial velocity within each grouping.
Positional rms uncertainties for the spots were estimated both from 
the position fit errors in each velocity plane and from the scatter in 
these values across the velocity width of the feature, demonstrating 
that the uncertainties are generally less than 0.025 arcsec.

The overall spatial pattern of the spots is displayed in Fig. 2.  
Note that 1 second  = 7.12 arcsec at this declination. 
Many coincidences occur between features of different polarization, and 
between the transitions.  For the purposes of position comparison 
between features, we treat any offset of up to about one-third of 
the halfpower beamwidth as a coincidence.  
Position coincidences of features with different polarization, and/or 
different transition, result in the 59 spectral features being located in 
just 28 different spots, as described below.

\begin{table*}

\caption{Polarized features of OH 300.969+1.147; position 
offsets relative to (J2000) R.A. = $12^h$ $34^m$ $53^s.27$, Dec = 
$-61^{\circ}$ $39'$$39''.63$. Details in section 4.1} 
\begin{center}

\begin{tabular}{ccrrrrr@{\hspace{1cm}}l}
  \hline
  Transition & Feature & Velocity & R.A.offset & Dec. offset & Peak flux & $B$ & Comment \\
  and pol. & label & (km s$^{-1}$) & (arcsec) & (arcsec) & (Jy beam$^{-1}$) & (mG) & \\
  \hline
\emph{1665LHC}  &  a    &  -34.69 &  -0.171 &    -0.045 &    0.11 & -1.1   & $Z_{1}$   \\
& b    &  -34.97 &   0.016 &    -0.004 &    2.73 & -1.3	& $Z_{2}$; matches $Z_{13}$ \\
& c    &  -35.71 &  -0.305 &    -0.574 &    1.16 & -4.2   & $Z_{3}$	  \\
& d    &  -35.86 &  -0.119 &     0.051 &    0.14 & -4.7 	& $Z_{4}$	  \\
& e    &  -36.19 &  -0.169 &    -0.457 &    0.16 & -4.6 	& $Z_{5}$	  \\
& f    &  -36.53 &  -0.195 &    -0.653 &    0.36 & -4.5 	& $Z_{6}$	  \\
& g    &  -36.78 &  -0.271 &    -0.549 &    0.67 & -3.2 	& $Z_{7}$	  \\
& h    &  -37.16 &  -0.283 &    -0.554 &    0.25 &      	& unpaired	  \\
& i    &  -37.21 &  -0.216 &    -0.664 &    0.27 & -4.0 	& $Z_{8}$; matches $Z_{15}$ \\
& j    &  -37.92 &  -0.175 &    -0.654 &    2.80 & -3.7 	& $Z_{9}$; matches $Z_{16}$ \\
& (k)    &  -38.15 &  -0.285 &    -0.562 &    0.10 &      	& (elliptical with F)	  \\ 
& (l)    &  -38.90 &  -0.168 &    -0.457 &    0.20 &      	& (elliptical with J)	  \\
& (m)    &  -39.20 &  -0.196 &    -0.652 &    0.13 &      	& (elliptical with K)	  \\
& n    &  -39.32 &   0.510 &    -0.560 &    0.30 & -3.2 	& $Z_{10}$  \\
& (o)    &  -40.05 &  -0.168 &    -0.652 &    0.14 &      	& (elliptical with M)	  \\
& p    &  -40.17 &   0.403 &    -1.465 &    1.71 & -4.4 	& elliptical with N; $Z_{11}$  \\
& (q)    &  -40.28 &   0.062 &    -0.305 &    0.27 &      	& (elliptical with P)	  \\
& (r)    &  -40.86 &   0.442 &    -0.529 &    0.16 &      	& (elliptical with R)	  \\
& s    &  -41.01 &   0.249 &    -1.502 &    0.21 & -1.5 	& $Z_{12}$; almost matches $Z_{17}$ \\ 
& (t)    &  -41.98 &   0.279 &    -1.498 &    0.06 &      	& (elliptical with T)	  \\
\emph{1665RHC} & A    &  -34.71 &  -0.125 &    -0.044 &    0.10 &      	& unpaired  \\ 
& B    &  -35.31 &  -0.164 &    -0.043 &    0.14 &      	& $Z_{1}$  \\	 
& C    &  -35.73 &   0.007 &    -0.001 &    1.64 &      	& $Z_{2}$  \\	 
& D    &  -37.30 &  -0.132 &     0.049 &    0.27 &      	& unpaired  \\	 
& E    &  -37.98 &  -0.123 &     0.046 &    0.15 &      	& unpaired  \\	 
& F    &  -38.16 &  -0.275 &    -0.558 &    0.40 &      	& elliptical with k; $Z_{3}$   \\	 
& G    &  -38.30 &  -0.127 &     0.065 &    0.09 &      	& unpaired  \\	 
& H    &  -38.66 &  -0.110 &     0.056 &    0.13 &      	& $Z_{4}$   \\	 
& I    &  -38.65 &  -0.259 &    -0.552 &    0.39 &      	& $Z_{7}$   \\	 
& J    &  -38.93 &  -0.165 &    -0.452 &    0.78 &      	& elliptical with l; $Z_{5}$  \\	 
& K    &  -39.18 &  -0.197 &    -0.653 &    5.40 &      	& elliptical with m; $Z_{6}$  \\	 
& L    &  -39.59 &  -0.220 &    -0.671 &   15.44 &      	& $Z_{8}$  \\	 
& M    &  -40.11 &  -0.177 &    -0.652 &    5.52 &      	& elliptical with o; $Z_{9}$  \\	 
& (N)    &  -40.16 &   0.405 &    -1.461 &    0.70 &      	& (elliptical with p)  \\	 
& O    &  -40.16 &  -0.508 &     0.144 &    0.50 &      	& unpaired  \\	 
& P    &  -40.42 &   0.053 &    -0.300 &    0.48 &      	& elliptical with q; unpaired  \\	 
& Q    &  -40.40 &  -0.083 &    -0.415 &    0.17 &      	& unpaired  \\	 
& R    &  -40.90 &   0.454 &    -0.534 &    0.20 &      	& elliptical with r; unpaired  \\	 
& S    &  -41.19 &   0.508 &    -0.557 &    0.43 &      	& $Z_{10}$  \\	 
& T    &  -41.89 &   0.292 &    -1.497 &    0.40 &      	& elliptical with t; $Z_{12}$  \\	 
& U    &  -42.59 &  -0.163 &    -1.378 &    0.16 &      	& unpaired \\	 
& V    &  -42.78 &   0.439 &    -1.504 &    0.57 &        & $Z_{11}$  \\                       
\emph{1667LHC} & a    &  -35.22 &   0.013 &    -0.003 &    2.96 & -1.2 	& $Z_{13}$; matches $Z_{2}$   \\	 
& b    &  -35.63 &  -0.104 &    -0.018 &    0.18 &      	& unpaired  \\	 
& c    &  -36.12 &  -0.036 &     0.031 &    0.08 & -1.3 	& $Z_{14}$  \\	 
& d    &  -36.24 &  -0.275 &    -0.740 &    0.12 &      	& unpaired  \\	 
& e    &  -37.09 &  -0.384 &    -0.602 &    0.06 &      	& unpaired  \\	 
& f    &  -37.73 &  -0.229 &    -0.670 &    0.57 & -3.5 	& $Z_{15}$; matches $Z_{8}$  \\	 
& g    &  -38.48 &  -0.176 &    -0.654 &    1.38 & -3.6 	& $Z_{16}$; matches $Z_{9}$  \\	 
& (h)    &  -39.86 &  -0.184 &    -0.648 &    0.10 &      	& (elliptical with D)  \\	 
& (i)    &  -40.25 &   0.052 &    -0.300 &    0.28 &      	& (elliptical with E) \\	 
& j    &  -40.54 &   0.254 &    -1.491 &    1.04 & -2.0 	& $Z_{17}$; almost matches $Z_{12}$   \\	 
& k    &  -41.04 &   0.458 &    -1.513 &    0.35 &      	& unpaired  \\	 
\emph{1667RHC} & A    &  -35.64 &   0.000 &     0.000 &    4.02 &      	& $Z_{13}$   \\	 
& B    &  -36.59 &  -0.021 &     0.043 &    0.13 &      	& $Z_{14}$  \\	 
& C    &  -38.97 &  -0.224 &    -0.672 &    4.34 &      	& $Z_{15}$  \\	 
& D    &  -39.74 &  -0.178 &    -0.651 &    0.37 &      	& elliptical with h; $Z_{16}$  \\	 
& E    &  -40.38 &   0.052 &    -0.299 &    0.35 &      	& elliptical with i; unpaired  \\	 
& F    &  -41.25 &   0.260 &    -1.490 &    0.35 &      	& $Z_{17}$  \\	 
  \hline					      
  \end{tabular}					      
  \label{table-lrpos}			      
\end{center}					      
\hspace{18mm}					      
\end{table*}

Ten features of Table 1 have a matching stronger feature in the opposite 
polarization but at the same position and velocity. For example, 1665 MHz 
LHC feature q matches stronger RHC feature P. In many similar cases 
studied 
in earlier work (e.g. section 4.3 of Fish \& Reid 2006), detailed 
investigation shows such features to be elliptically polarized, sometimes 
with additional unpolarized emission.  Our data, without linear 
polarization analysis, do not distinguish these situations, but for 
simplicity, we add the comment 
to q: `elliptical with P', and likewise the comment `elliptical with q' is 
added to P.  The weaker component is not a truly separate feature, and is 
distinguished in Table 1 by adding parentheses to both its 
feature label in column 1 and its comment.  
This particular example is of additional interest because it is not part 
of a recognisable Zeeman pair, and it is co-located with an 
unpaired 1667-MHz LHC feature i (similarly possessing a matching stronger 
feature (E) in RHC polarization).

Spatially coincident features of different circular polarization which 
have different velocities are expected to be the $\sigma$-components of a 
Zeeman pattern, The $\sigma$-components are most generally elliptically 
polarized, usually with the circular percentage dominating over the 
linear percentage.  

We interpret 34 features as components of 17 Zeeman pairs (12 at the 
1665-MHz transition and 5 at 1667 MHz), as noted in 
the comments column of Table 1;  the 
derived magnetic fields are given in the previous column. 
The magnetic field values assume Zeeman splitting 
factors such that, in a 1 mG magnetic field, a spectral
line will be split into two components of opposite circular polarization,
separated in frequency by an amount equivalent to 0.590 \kms\ for the
1665-MHz transition and 0.354 \kms\ for the 1667-MHz transition.

Three of the 
1667-MHz Zeeman pairs closely match corresponding 1665-MHz Zeeman pairs, 
as noted in the comments.  Zeeman pairs Z12 at 1665 MHz and Z17 at 1667 
MHz are co-located, and match in magnetic field, but differ by 0.5 \kms\  
in their mean velocity.  In addition to the 14 different spot 
positions hosting Zeeman pairs, there are 14 spot positions (one of 
them with emission at both 1665 and 1667 MHz) which 
have `unpaired' spectral features i.e. features that are not part of a 
Zeeman pair.

Further notes on some of the individual spectral features are given below; 
these include consideration of single dish Parkes spectropolarimetry 2005 
November, which obtained full polarimetry at 1665 and 1667 MHz with high 
spectral resolution but low spatial resolution.
\\
\\
\subparagraph{1665 MHz LHC g.} 
Single dish spectra from the Parkes telescope suggest 35 per 
cent linear polarization of this feature.  We note that the Parkes 
observations were several years later than the LBA observations, and this 
feature in particular had increased in intensity, but it is useful to 
explore the possible implications of the Parkes data for interpreting the 
LBA data.  If 35 per cent linear polarization is present in the LBA data, 
then the maximum possible 
circular polarization would be 94 per cent, and there would then be an 
expectation of a LHC component of 0.05 Jy. Since this is close to our 
detection limit, our non-detection of a LHCP feature is not significant.  
We conclude that feature g is most likely elliptically polarized, with the 
RH circular component dominant.  

\subparagraph{1665 MHz LHC m.}
Coincidence with feature K is noted as evidence for linear polarization in 
K.  This is supported by the single dish data from Parkes 
indicating at least 10 per cent linear polarization of this feature.

\subparagraph{1665 MHz LHC p.}
Commented as `elliptical with N'.  This is supported by single dish 
spectra 
from the Parkes telescope showing at least 10 per cent linear polarization 
at this velocity.  The circular polarization is 71 per cent and so if the 
feature is 100 per cent polarized, then its linear polarization would also 
be 71 per cent.  

\subparagraph{1665 MHz RHC L.}
Single dish spectra from the Parkes telescope suggest 10 per 
cent linear polarization at this velocity.  The maximum possible circular 
polarization would then be 99.5 per cent, and there would then be an 
expectation of a LHC component of 0.077 Jy.  Although this is not 
detected, the absence may result from some maser variability in the 
several years between LBA and Parkes observations. 
We conclude that, although the RH circular component is dominant, there is 
likely to be some linear polarization also present, to give net 
elliptical polarization.

\subparagraph{1667 MHz RHC C.}
Single dish spectra from the Parkes telescope suggest 10 per linear 
polarization at this velocity.  The maximum possible circular polarization 
would then be 99.5 per cent and the expectation of a 20 mJy LHC component.  
This is below our sensitivity level, so its absence is not 
significant.
\\

\subsection{Velocities, magnetic field, and kinematics}

From the median velocity of detected maser features, we can estimate the
systemic velocity of OH 300.969+1.147 to be -38 \kms.  The \HII\ region in 
this direction has a radial velocity of -47 \kms\ (Caswell \& Haynes 
1987b).  
The `tangent point' velocity, the most negative `allowed' velocity in this 
direction for commonly used  Galactic rotation models, is -40 \kms.  We 
conclude that the maser and \HII\ region are close to the tangent point, 
and thus at a uniquely defined distance of 4.3 kpc (if the Galactic
Centre is assumed to be at 8.5 kpc from the Sun; or  5.1 kpc distant if 
the Galactic Centre were at 10 kpc, the value assumed in older studies).

Considering the Zeeman pairs in more detail, we note that, throughout the 
whole site, the
estimates of magnetic field for all 17 Zeeman pairs are notably in
the same sense (towards the observer, as denoted by a negative field) and 
range from -1.1 to  -4.7 mG, with a median value of -3.5 mG.

\begin{figure*}
\centerline{\epsfxsize=13cm\epsfbox{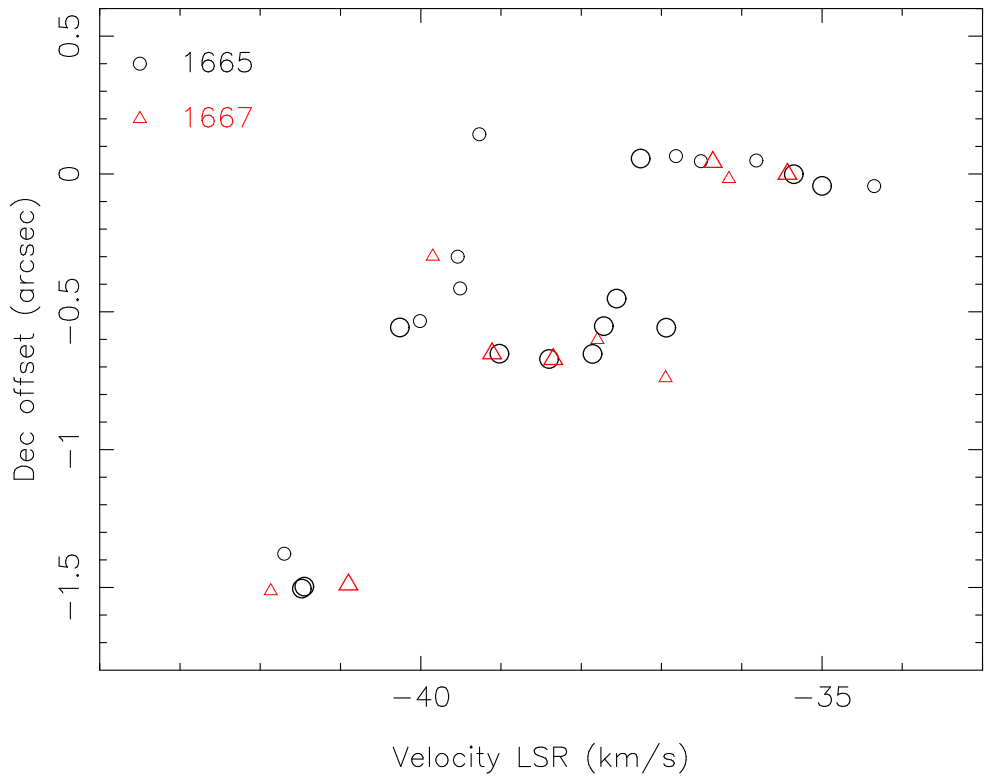}}
\caption{Declination of each maser spot is plotted against `demagnetized' 
velocity.  Each large symbol corresponds to a Zeeman pair with its mean 
velocity plotted, and each small symbol is an unpaired single 
feature with velocity adjusted as described in Section 4.2}
\end{figure*}

Further information on the magnetic field is provided by excited-state OH 
 6035- and 6030-MHz masers.  A position measurement 
for strong 6035-MHz emission at -37.5 \kms\ (Caswell 1997) 
yields a position RA 12$^h$34$^m$53.27$^s$, Dec. 
-61$^{\circ}$39${\arcmin}$39.9${\arcsec}$.  
The spectra in Fig. 1 of Caswell (2003) show that this 6035-MHz reference 
feature 
is a Zeeman pair arising in a magnetic field of -5.0 mG, and is 
accompanied by an OH 6030-MHz pair indicative of a similar field.  Two 
other weak 6035-MHz Zeeman pairs corroborate this field strength.  
The 6035-MHz position is shown on Fig. 2 as a dotted square;  the 
uncertainty in the position is larger than the square, with contributions 
from the absolute position registration for both species, but the location 
lies, encouragingly, amongst a high concentration of 1665- and 
1667-MHz maser spots of which several reveal similar magnetic fields.  
Indeed, it may be closely associated with the OH 1665-MHz Zeeman pair Z5 
which has essentially the same magnetic field (-4.6 mG), same mean 
velocity, and is coincident (the offset of 0.23 arcsec is smaller than the 
registration uncertainty).

The 6-GHz ATCA observations (Caswell 1997) also reveal a weak methanol 
6668-MHz maser close to the OH 6035-MHz maser, and superposed on a uc\HII\ 
region with flux density 164 mJy at 6 GHz.  Lower resolution maps of 
the region at 5 GHz reveal a much higher flux density of 3.1 Jy 
extending over an estimated size of 1 arcmin (Caswell \& Haynes 1987b), so 
the region is clearly quite complex, with additional emission on a much 
more diffuse scale than 1 arcsec. 
In the absence of full mapping of the uc\HII\ region, its boundary is 
sketched as a large dotted circle on Fig. 2 circumscribing the maser spots 
and consistent with its absolute position measurement.  

We now explore the kinematics, and whether the velocity field, as derived 
from the combination 
of the spatial distribution of the maser spots and their radial 
velocities, might reveal a pattern indicative of rotation or expansion.  

However, we ought first to correct for the fact that velocity values at 
1665 and 1667 MHz are not purely kinematic and have been significantly 
affected
by the magnetic field.  In order to obtain a corrected velocity estimate
at each position, we first use the mean velocity value for each of the 
17 Zeeman pairs.   The corrections relative to individual spot velocities 
range from 0.18 \kms\ for the lowest-field 1667-MHz feature to nearly 1.5 
\kms\ for the highest-field 1665-MHz feature.   For the other (15) 
unpaired features, we must first decide whether they are more likely to be 
 $\pi$-components or lone surviving $\sigma$-components from Zeeman 
patterns.  We note that none of them has location and velocity compatible 
with being the $\pi$-component of any of our recognized Zeeman pairs, and 
regard them as most likely lone surviving $\sigma$-components.  We 
`correct' their velocities by assuming that they originated in the same 
field as their nearest Zeeman pair companion.  All `corrected' velocities 
were then inspected over the whole map and the only suggestive pattern was 
an apparent gradient of velocity with declination.  
A plot of velocity against declination was then made, and is shown in Fig. 
3.  Large symbols are used for the Zeeman patterns 
since they represent the information from 2 spots rather than 1, and have 
more reliable corrections.  Fig. 3 reveals a tendency for more 
negative velocities to be at the south of the distribution, with gradient 
of approximately 4 \kms\ over 1 arcsec.  
Naively we could regard this as indicating rotation about an east-west 
axis.  

\subsection{Properties of the site as a whole}

With regard to other OH transitions, we note the detection of a loosely 
associated 1720-MHz maser (Caswell 2004) at (J2000) RA 
12$^h$34$^m$53.63$^s$, Dec. -61$^{\circ}$39${\arcmin}$40.0${\arcsec}$) 
with uncertainty 0.4".  It lies outside the region shown in Fig. 2, with 
an offset to larger RA by 0.36s (= 2.6") and south by 0.4" from the 
reference location used in this study (for Table 1 and Fig. 2).  
The 1720-MHz maser displays a 
clear Zeeman pattern indicative of a magnetic field of -5 mG. with mean 
velocity -42.5 \kms.  It lies in a broad absorption dip extending from 
-39 to -47 \kms, presumably indicative of a dense,  more extensive, host 
molecular cloud.
The spatial offset of the 1720-MHz maser from our current object of study 
seems large enough to rule out a common source of excitation, but it 
clearly lies in the same region of massive star formation.  This is 
corroborated by its radial velocity and its magnetic field value, 
both of which are in the same range as our 1665 and 1667-MHz measurements.  

There has been no detection of OH at the 1612-MHz transition (Caswell 
1999). 

In addition to the presence of the previously mentioned methanol maser at 
the 6668-MHz transition, there is a methanol maser at the 12-GHz
transition with a weak peak at the same velocity (Caswell et al. 1995).  

Water maser emission peaking above 40 Jy and covering the velocity 
range -43 to -86 \kms\ (Caswell et al. 1989) was reported at approximately 
this position but is now found (Caswell, ATCA unpublished data) to be 
offset 18 arcsec, and thus not closely related.  

The general location of the star formation complex is at a Galactic 
latitude of more than 1$^{\circ}$ where, with 301.109+0.969 (Caswell 
\& Haynes 1987b), it is one of only two prominent star 
forming regions in this portion of sky.  Their separation of 
0.246$^{\circ}$ corresponds to 18 pc.  This comparative isolation will 
allow detailed future studies to be undertaken unaffected by other nearby 
companion star clusters, an isolation which 
can be especially valuable when investigating the more common molecular 
species that are able to trace the properties of a massive-star-forming  
environment.  

Summarizing the properties, we conclude that the region, although young 
enough to host masers of OH and methanol, is mature enough to have a 
prominent uc\HII\ region.  The relative weakness of methanol, plus the 
absence of an intimately associated water maser, are also common 
properties of these more mature regions (Caswell 1997).  The complex as a 
whole does, however, also 
host a water maser, a 1720-MHz OH maser, and a more extensive OH cloud 
(responsible for a prominent feature of 1720 MHz absorption against the 
extended \HII\ region).  The 
consistency of magnetic field direction within the source (and even at the 
location of 1720-MHz emission) also reinforces its status as a stable 
region which is approaching the end of the short evolutionary period 
that can support maser emission, according to current ideas (Caswell \& 
Reynolds 2001).  

\subsection{Comparison of OH 300.969+1.147 with other maser sites}

The most recent major addition to high resolution mapping of OH masers 
is from a study of 18 targets with the VLBA by Fish, Reid, Argon \& 
Zheng (2005), discussed in detail by Fish \& Reid (2006).  
Very few targets conform precisely to the canonical picture, in which 
maser spots appear projected onto a simple uc\HII\ region, with magnetic 
field estimates all of the same sign, and some kinematic order 
discernible.  

Most of the VLBA targets lie in 
the Galactic first quadrant, longitude 0 to 90$^{\circ}$, and the only 
one, 351.775-0.536 that lies in the fourth quadrant (longitude 270 to 
360$^{\circ}$) is not covered over its full velocity range. 

The present observations add another example from the largely untapped 
resource of southern fourth quadrant masers and establishes another 
simple morphology against which we can compare the more pathological ones.  

In a previous study (Caswell \& Reynolds 2001), we compared the properties 
of OH 323.459-0.079 with those of the archetypical W3(OH), and concluded 
that it was similar, but slightly larger in linear extent and weaker, and 
thus perhaps slightly older.  The present target also appears larger 
(largest separation of maser spots is 36 mpc), weaker, and it has 
relatively weaker methanol maser emission and no water maser.  All of 
these slight differences conform to expectations for a slightly more 
evolved maser site.  

Maser sites show a weak tendency to display a magnetic field direction in 
the sense of Galactic rotation, and indeed the present target conforms to 
this, with a field towards us in the Galactic fourth quadrant.  Some 
counter-examples such as OH 34.257+0.154 occur in crowded and apparently 
disturbed regions of the Galactic plane which may be relevant.  
However, OH 323.459-0.079 shows a magnetic field counter to Galactic 
rotation, and 
yet lies in an apparently undisturbed region.  Indeed, it lies at similar 
distance, and in the same (Crux-Scutum) spiral arm as OH 300.969+0.147,  
with its main point of difference being its location at the inner edge of 
the arm rather than the outer edge.  Another notable counter-example is 
285.263-0.050 which appears to lie close to the centre of the Carina arm.  
Thus at present, there is no evidence that the counter-examples are merely 
exceptions to a general rule because of their exceptional circumstances.  
Rather, the relation appears somewhat weak in its significance.  

\section[]{Conclusion}

In the common picture of OH masers in massive young stellar objects, a 
cluster of maser spots is projected against an uc\HII\ region of 
diameter about 30 mpc, with the pattern of maser spots revealing a 
magnetic field of a few mG, and a small velocity range of less than 10 
\kms, with velocity field perhaps indicative of rotation.  The 
detailed study of W3(OH) has defined this archetype, but 
finding additional real examples conforming to this pattern has proved 
difficult.   The present study establishes OH 300.969+1.147 as a new 
example conforming to this simple pattern, and we regard it as a mature 
region nearing the end of the brief evolutionary 
period that can support maser emission.  
Examples of these canonical maser sites are important for comparison with 
other diverse sites that have no discernible uc\HII\ region (perhaps 
younger), or with maser spots scattered over a larger area (perhaps 
multiple sources), magnetic fields more complex, and velocity patterns 
suggestive of outflows.

\section*{Acknowledgments}

We thank the ATNF staff who make the intricate operation of the LBA
successful, and fellow 
astronomers who participated in the observing session.  
BHK thanks ATNF for hospitality and funding as a distinguished visitor 
during some of the period when this work was performed.

\bsp
\label{lastpage}
\end{document}